\documentclass[10pt]{amsart}

\usepackage{tikz}
\usetikzlibrary{positioning}
\usetikzlibrary{arrows,arrows.meta}
\usepackage{amssymb,amsmath,latexsym,graphicx}
\usepackage{charter,eucal}
\usepackage{amssymb}
\usepackage{amscd}
\usepackage{tcolorbox}
\usepackage[all,cmtip]{xy}
\usepackage{rotating,multicol,soul}
\usetikzlibrary{decorations.markings}

\setul{}{1.1pt}

\newtheorem{theorem}{Theorem }[section]

\newtheorem{lemma}[theorem]{Lemma}
\newtheorem{observation}[theorem]{Observation}

\newtheorem{remark}[theorem]{Remark}
\newtheorem{corollary}[theorem]{Corollary}
\newtheorem{proposition}[theorem]{Proposition}
\newtheorem{problem}[theorem]{Problem}
\newtheorem{question}[theorem]{QUESTION}
\newtheorem{principle}[theorem]{\textsc{Principle}}

\newcommand{\bt}{\begin{theorem}}
\newcommand{\et}{\end{theorem}}
\newcommand{\bmt}{\begin{maintheorem}}
\newcommand{\emt}{\end{maintheorem}}
\newcommand{\bc}{\begin{corollary}}
\newcommand{\bl}{\begin{lemma}}
\newcommand{\ec}{\end{corollary}}
\newcommand{\el}{\end{lemma}}
\newcommand{\bo}{\begin{observation}}
\newcommand{\eo}{\end{observation}}
\newcommand{\bp}{\begin{proposition}}
\newcommand{\ep}{\end{proposition}}
\newcommand{\br}{\begin{remark}}
\newcommand{\er}{\end{remark}}
\newcommand{\bpr}{\begin{principle}}
\newcommand{\epr}{\end{principle}}
\newcommand{\bq}{\begin{question}}
\newcommand{\eq}{\end{question}}
\newcommand{\bpro}{\begin{problem}}
\newcommand{\epro}{\end{problem}}

\def\Aut{\mathrm{Aut}}

\def\C{\mathbb{C}}

\def\mR{\mathcal{R}}

\def\eop{\hspace*{\fill}$\blacksquare$}

\def\id{\mathrm{id}}

\def\R{\mathbb{R}}

\newcommand{\F}{\mathbb{F}}
\newcommand{\bP}{\mathbb{P}}
\newcommand{\mC}{\mathcal{C}}

\newcommand{\mX}{\mathcal{X}}

\newcommand{\mL}{\mathcal{L}}

\newcommand{\mF}{\mathcal{F}}

\newcommand{\mT}{\mathcal{T}}

\newcommand{\mB}{\mathcal{B}}
\newcommand{\mH}{\mathcal{H}}

\usetikzlibrary{matrix}
\usetikzlibrary{arrows}
\usepackage{pgf}
\usepackage{lscape}
\usepackage{listings}

\usepackage{enumitem}

\title{Quantum theory without the Axiom of choice, and Lefschetz Quantum Physics}
\author{Koen Thas}
\thanks{}
\address{Ghent University, Department of Mathematics: Algebra and Geometry, Krijgslaan 281, S25, B-9000, Ghent, Belgium}
\email{koen.thas@gmail.com}\date{}

\begin{document}

\maketitle
\begin{abstract}
In this conceptual paper, we discuss quantum formalisms which do not use the famous Axiom of Choice. We also consider the fundamental problem which addresses the (in)correctness of having the complex numbers as the base field for Hilbert spaces in the K\o benhavn interpretation of quantum theory, and propose a new 
approach to this problem (based on the Lefschetz principle).  Rather than a Theorem--Proof--paper, this 
paper describes two new research programs on the foundational level, and focuses on fundamental open questions in these programs which come along the way.  \\
\end{abstract}

\begin{tcolorbox}
\tableofcontents
\end{tcolorbox}

\section{Introduction}

The K\o benhavn interpretation over the complex number has proved its uncanny effectiveness since it was conceived almost a century ago. On the algebraic level, states are solutions of a Schr\"{o}dinger-type first order differential equation such as  
\begin{equation}
\label{schro}
\frac{\partial{\vert \psi(t) \rangle}}{\partial{t}} = \widehat{H}\Big(\vert \psi(t) \rangle\Big),
\end{equation}
and linear combinations of solutions are also solutions. Equations such as (\ref{schro}) were developed in a complex model, given it an assumed complex nature. 
On the other hand, the superposition principle  for linear combinations works perfectly over any field, and equations such as (\ref{schro}) | or rather, the physical behavior they describe  | can also be developed over other fields. The set of all solutions form a vector space over a commutative field or more generally, over a division ring.  

The fact that every polynomial in one variable with complex coefficients has a complex root yields the essential property that  square complex matrixes always have eigenvalues, and  on the level of observables (see section \ref{KOB}), this is obviously a very important property. Also, because the subfield $\mathbb{R}$ is of degree $2$, the inner product used in the K\o benhavn description (cf. section \ref{KOB}) allows one to have a normalization process and probability theory at one's disposal. 

Finally, elegance and simplicity of the model adds to its value. In fact, we are convinced that this side of the story, together with the previous point, forms the main ``obstacle'' in considering other K\o benhavn models over different (algebraically closed) fields, or even more general algebraic structures. In mathematics for instance, for a very long time algebraic geometers solely systematically worked over the complex numbers, due to the very same reasons quantum physicists work virtually solely over $\C$: for its beauty and effectiveness (algebraically closed) and simply, because it works (so why even bother?). Only when people such as Andr\'{e} Weil  started to consider and study algebraic geometry over finite fields   in a systematic way, it turned out that this brave new algebraic geometry was as interesting (and arguable even more interesting) than the ``old'' theory, and eventually even added much to the understanding of complex algebraic geometry.

This very discussion certainly is one of the leading threads in this paper, and motivates us to introduce a conjectural Lefschetz quantum principle below. 

Secondly, {\em if} we assume to work over $\C$, then often implicitly the infamous Axiom of Choice is used in the mathematical machinery used to describe quantum physics. But since the philosophical discussion underlying quantum physics | including its many famous thought experiments | is extremely important for the theory, and since mathematics | even basic linear algebra | is so different without the Axiom of Choice, we want to investigate what happens to quantum theory if one does not rely on the Axiom of Choice.       

{\footnotesize We refer to the essay \cite{JBKT} for a deeper philosophical discussion on this matter.}

\subsection*{Plan of the present paper}

In section \ref{AOC}, we will mathematically discuss the Axiom of Choice. In section \ref{KOB}, we tersely describe the author's approach to general quantum theories (over division rings) and we pay particular attention to finite fields and the minimal model. 
 In section \ref{Lefsch}, we develop a conjectural Lefschetz principle for quantum theory, and to that end we first mathematically discuss the classical Lefschetz principle. The next section is the short section \ref{autcode}, in which we develop some basic mechanisms to develop quantum codes in models with the Axiom of Choice. In the final section, we discuss the impact of not accepting the Axiom of Choice on measurements and probabilities, and devise a number of thought experiments.

\subsection*{Acknowledgments}

The author wishes to thank Karl Svozil and Bill Wootters for a number of interesting and helpful communications.

\section{Axiom of Choice: mathematical discussion}
\label{AOC}

We start this section with a first formal formulation of the Axiom of Choice (AC):
\begin{tcolorbox}
\begin{quote}
``For every indexed family ${(S_i)}_{i\in I}$ of nonempty sets, there exists an indexed family ${(x_i)}_{i \in I}$ such that $x_j \in S_j$ for every $j \in I$.''  
\end{quote}
\end{tcolorbox}

Let us look at a first illuminating example. Suppose each $S_i$ is a subset of the positive integers $\mathbb{N}$; then we could define $x_i$ as the smallest number in $S_i$. The function which assigns to each $S_j$ the element $x_j$ 
is called a {\em choice function}. In this case, we do not need to invoke the Axiom of Choice in order to fulfil the desired property. But in case we define $\{ S_i \ \vert\ i \in I \}$ as the set of nonempty subsets of the reals $\R$, no such a choice function is known. 

\subsection{Russel's socks}

Bertrand Russel gave a particularly entertaining example where one has to invoke the Axiom of Choice in order to define a choice function. Suppose one has an infinite collection of pairs of socks, where we assume that in one pair of socks there is no way to distinguish between the socks. Then in order to select one sock from each pair, we need to invoke AC. Note that if we had started from pairs of shoes instead of socks, we {\em could} have defined a choice function (``select the left shoe''). 

\subsection{Choice functions and Cartesian products}

In terms of choice functions, we can formulate AC as follows:
\begin{tcolorbox}
\begin{quote}
``For any set $\mX$ of nonempty sets, there exists a choice function $f$ defined on $\mX$ which maps each set of $\mX$ to an element of that set.''  
\end{quote}
\end{tcolorbox}

Obviously, each such choice function $f$ defines an element of the Cartesian product of the sets in $\mX$, so that we can give another equivalent statement:
\begin{tcolorbox}
\begin{quote}
``For any set $\mX$ of nonempty sets, the Cartesian product of the sets in $\mX$ is nonempty.''  
\end{quote}
\end{tcolorbox}

\subsection{Vector spaces}

The equivalent formulation which interests us the most in the context of the present paper, is the following.
\begin{tcolorbox}
\begin{quote}
``Every vector space has a basis.''  
\end{quote}
\end{tcolorbox}

Obviously, in the context of quantum theory such results are extremely important! 

Note that upon accepting the Axiom of Choice, one can show that the size of a basis of a given vector space is independent of the choice of basis, and this size yields a notion of {\em dimension}.

\section{K\o benhavn interpretations beyond $\C$}
\label{KOB}

In classical quantum theory following the K\o benhavn interpretation | in some papers called ``Actual Quantum Theory'' (AQT) | the state space is a Hilbert space (foreseen with the standard inner product). More precisely:
$\left\{
\begin{tabular}{p{.9\textwidth}}
\begin{itemize}
\item[(HS)]
a physical quantum system is represented by a Hilbert space $\mH = \Big((\mathbb{C}^\omega,+,\cdot),\langle \cdot,\cdot \rangle\Big)$, with $\langle \cdot,\cdot \rangle$ the standard inner product and $\omega$ allowed to be non-finite;
\item[(IP)]
the standard inner product $\langle \cdot,\cdot \rangle$ sends $\Big((x_1,\ldots,x_{\omega}),(y_1,\ldots,y_{\omega})\Big)$ to $\overline{x_1}y_1 + \cdots + \overline{x_{\omega}}y_{\omega}$ (or $x_1\overline{y_1} + \cdots + x_{\omega}\overline{y_{\omega}}$), where $\overline{c}$ is the complex conjugate of $c \in \mathbb{C}$; complex conjugation is an involutary automorphism of the field $\mathbb{C}$; 
\item[(PS)]
up to complex scalars, pure states (wave functions) are represented by nonzero vectors in $\mathbb{C}^\omega$; usually, one considers normalized vectors; 
\item[(TE)]
time evolution operators are represented by linear operators of $\mathbb{C}^\omega$ that preserve $\langle \cdot,\cdot \rangle$, that is, {\em unitary operators}. If $\omega$ is finite, unitary operators correspond to nonsingular complex $(\omega \times \omega)$-matrices $U$ such that $UU^* = \id$; 
\item[(OB)]
measuring an observable $A$ in a system described by the wave function $\vert \psi \rangle$, amounts to collapsing $\vert \psi \rangle$ into one of the orthogonal eigenvectors $\vert \psi_i \rangle$ of the Hermitian operator $A$, yielding as measurement the corresponding eigenvalue $\lambda_i$; 
\item[(TP)]
composite product states correspond to tensor products $\vert \psi_1 \rangle \otimes \vert \psi_2 \rangle \in \mH_1 \otimes \mH_2$; if a state in $\mH_1 \otimes \mH_2$ 
is not a product state, it is entangled;
\item[(BR)]
one follows Born's rule, which says that $\vert \langle \psi, \psi_i \rangle \vert^2$ is the probability that the measurement $\lambda_i$ will be made;
\item[(XX)]
($\cdots$). 
\end{itemize}
\end{tabular}
\right\}$

{

In the rest of this section we tersely explain our approach of \cite{GQT}, which unifies all known modal quantum theories (over finite fields $\F_q$, algebraically closed fields, general division rings with involution). 

\subsection{The general setting}

A {\em division ring} is a field for which the multiplication is not necessarily commutative. Sometimes they are called ``skew fields,'' but we prefer the 
name division ring. In \cite{GQT} we described a general K\o benhavn approach in which all known classical and modal quantum theories are united in one and the same 
framework. The main philosophy is that instead of the field of complex number or finite fields, the underlying coordinatizing structures are generalized 
to division rings, so that we consider generalized Hilbert spaces over division rings instead of complex Hilbert spaces or their finite field analogons. Of course, one has 
to have a good alternative for the classical inproducts, and this is perfectly possible if we foresee the division rings with a so-called {\em involution} (cf. the next subsection). 
The details can be found in the next subsection.

\subsection{Standard $(\sigma,1)$-Hermitian forms}

A ``division ring with involution'' is a division ring with an involutory anti-automorphism.
If $k$ is a division ring with involution $\sigma$, the {\em standard $(\sigma,1)$-Hermitian form} on the right vector space $V(d,k)$, is given by
\begin{equation}
\Big\langle x,y \Big\rangle\ :=\ x_1^\sigma y_1 + \cdots + x_d^\sigma y_d, 
\end{equation}
where $x = (x_1,\ldots,x_d)$ and $y = (y_1,\ldots,y_d)$. 

\begin{remark}{\rm 
In the case that $\sigma = \id$, we obtain a form which is usually called {\em symmetric}; it is not a proper Hermitian form, but still comes in handy in some situations
(for example in cases of field reduction: ``real Hilbert spaces'' have often been considered in quantum theory; see e.g. the work of Wootters et al. \cite{Wootreal,Wootreal3}).}
\end{remark}

 We propose to describe all classical and modal quantum theories in one and the same 
framework, under the umbrella of ``General Quantum Theories'' (GQTs), as follows:\\ 

\begin{center}
$\left\{
\begin{tabular}{p{.9\textwidth}}
From now on, we propose to depict a physical quantum system in a \ul{general Hilbert space $\mH = \Big((V(\omega,k),+,\cdot),\langle \cdot,\cdot \rangle\Big)$, with $k$ a division ring with involution $\sigma$, and $\Big\langle \cdot,\cdot \Big\rangle$ a $(\sigma,1)$-Hermitian form.} \\
\end{tabular}
\right\}$
\end{center} 

\bigskip
Following \cite{GQT}, we speak of a  {\em standard GQT} if  given an involution $\sigma$, the general Hilbert space comes with the standard $(\sigma,1)$-Hermitian form. As some fields such as 
the real numbers and the rational numbers do not admit nontrivial involutions, they only can describe ``improper'' quantum systems. 

\subsection{Algebraically closed fields}

Let $k$ be any algebraically closed field in characteristic $0$. 
It is well known that upon accepting the Axiom of Choice, there exists an involution $\gamma$ in $\Aut(k)^\times$, where $\Aut(k)$ denotes the automorphism group of $k$. 
Now consider the set
\begin{equation}
k_{\gamma} := \{ \kappa \in k\ \vert\ \kappa^{\gamma} = \kappa \}. 
\end{equation}

One easily shows that $k_{\gamma}$, endowed with the addition and multiplication coming from $k$, is also a field. There is however more by \cite{ArtSch}:

\begin{tcolorbox}
\begin{theorem}[$(\mathbb{C},\mathbb{R})$-Analogy | algebraically closed version in char. $0$]
\label{analogy}
Let $k$ be any algebraically closed field in characteristic $0$. Let $\gamma$ be an involution in $\Aut(k)^\times$. Then $-1$ 
is not a square in $k_{\gamma}$. Suppose $i \in k$ is such that $i^2 = -1$. Then $k = k_{\gamma} + i\cdot k_{\gamma}$
and $[k : k_{\gamma}] = 2$.  
\end{theorem}

So each element of $k$ has a unique representation as $a + bi$, with $a, b \in k_{\gamma}$ and $i$ a fixed solution of $x^2 = -1$. Fields which have index $2$ in their
algebraic closure are called {\em real-closed fields}, and can always be constructed as a $k_{\gamma}$ of some involution $\gamma$. Real-closed fields
share many properties with the reals $\mathbb{R}$: each such field is {\em elementarily equivalent} to the reals, which by definition means that {it has the 
same first-order properties as the reals.}  We call a GQT {\em complex-like} if it is defined over an algebraically closed field $k$ with nontrivial involution $\gamma$, 
where the elements of $k$ are represented in Theorem \ref{analogy} with respect to the field $k_{\gamma}$. 
\end{tcolorbox}

\begin{remark}{\rm 
The analogy goes even further: once we have defined $k_{\sigma}$ as above, and we represent each element $x$ in $k$ as $x = u + iv$ (which can be done by Theorem \ref{analogy}), it can be shown that the automorphism
$\sigma$ is given by
\begin{equation}
\sigma:\ k \mapsto k:\ u + iv \mapsto u - iv \ \ \ \mbox{(complex conjugation)}. 
\end{equation}
}\end{remark}

\subsection{Extension of quantum theories}
\label{ext}

If we consider a GQT $\mT$ over a field $k$ in characteristic $0$,  the following fundamental question arises: 
\begin{quote}
Is $\mT$ {\em embeddable}
in a complex-like theory (or in any other GQT, for that matter)? 
\end{quote}

Here, the notion of ``embeddable'' is obvious: if $k$ comes with the involution $\gamma$, we want to have a field extension $\ell \Big/ k$ for which  
$\ell$ is algebraically closed, and an involution $\overline{\gamma}$ of $\ell$ for which the restriction of $\overline{\gamma}$ to $k$ precisely is $\gamma$. 
Since any GQT
depends only on the Hermitian matrix of the $(\sigma,1)$-Hermitian form with respect to a chosen basis (with suitable adaptation to the infinite-dimensional case), 
it is clear that if the aforementioned GQT comes with matrix $A$ over $k$, then the same matrix $A$ defines a $(\overline{\gamma},1)$-Hermitian form over $\ell$ which 
induces the initial form over $k$.   

\begin{observation}
If we fix the dimension of the Hilbert space, then any GQT over $k$ (and with respect to $\gamma$) is part of the GQT over $\ell$ (with involution $\overline{\gamma}$). 
\end{observation}

The reason why a good extension theory is desired, is explained in the next paragraph.

\subsection*{COMPARISON THEORY}

Any two fields $k$ and $k'$ of the same characteristic are contained (as a subfield) in a field $\ell$. The following construction is simple: let $\wp$ be the prime field 
in both $k$ and $k'$ (isomorphic to $\mathbb{Q}$ in characteristic $0$ and to $\F_p$ in characteristic $p > 0$), generated by $0$ and $1$. Then put $k = \wp(S)$ and $
k' = \wp(S')$, with $S$ (resp. $S'$) a basis over $\wp$ of $k$ (resp. $k'$) consisting of algebraic and transcendental elements over $\wp$. Then $\wp(S \cup S')$ is 
``the'' desired field. Obviously such a field $\ell$ with the extension property is not 
unique, since $\ell$ can be extended indefinitely (for instance, by adding transcendental elements to $\ell$). If we have formulated a good extension formalism for general quantum theories (over algebraically closed fields), 
we would be able to evaluate problems formulated over $k$ and $k'$ in one and the same theory formulated over $\ell$ (and any of its extensions). In that way, if we fix the characteristic of the fields, we could look at a quantum theoretical setting prepared over different fields $k$ and $k'$ as being two guises of the same story: just extends 
the GQTs over $k$ and $k'$ to the appropriate GQT over $\ell$.     Since it is sometimes preferable to work over algebraically closed fields, we want essentially that the following 
diagram commutes, after the map $\mathsf{GQT}$ is applied which associates with each couple $(\rho,\phi)$ (where $\rho$ is a field or division ring, and $\phi$ an involution of $\rho$) the corresponding standard general quantum theory. (Of course, the same ideas can be expressed for non-standard GQTs as well.)

\begin{figure}[h]
\tikzset{every picture/.style={line width=0.5pt}} 

\begin{tikzpicture}[x=0.75pt,y=0.75pt,yscale=-0.75,xscale=1]

\draw [orange,-{>[scale=2.0]},line width=1.25]    (346,163) -- (443.17,256) ;
\draw [orange,-{>[scale=2.0]},line width=1.25]    (315,163) -- (222.89,256) ;
\draw [orange,-{>[scale=2.0]},line width=1.25]    (450,292) -- (450,420) ;
\draw [orange,-{>[scale=2.0]},line width=1.25]    (225,292) -- (225,420) ;
\draw [orange,-{>[scale=2.0]},line width=1.25]    (230,467) -- (319.32,565) ;
\draw [orange,-{>[scale=2.0]},line width=1.25]    (451,467) -- (365.51,565) ;

\draw (310,135.9) node [anchor=north west][inner sep=0.75pt]    {$\wp = \mathbb{Q} ,\mathbb{F}_{p} \ $};
\draw (188,260.9) node [anchor=north west][inner sep=0.75pt]    {$( k,\sigma _{k})$};
\draw (430,267.9) node [anchor=north west][inner sep=0.75pt]    {$( k',\sigma _{k'\ })$};
\draw (433,435) node [anchor=north west][inner sep=0.75pt]    {$\left(\overline{k'} ,\overline{\sigma _{k'}}\right)$};
\draw (190,435) node [anchor=north west][inner sep=0.75pt]    {$\left(\overline{k\ } ,\overline{\sigma _{k}}\right)$};
\draw (321,582.9) node [anchor=north west][inner sep=0.75pt]    {$( \ell ,\sigma _{\ell })$};

\end{tikzpicture}
\caption{Extension diagram. Each arrow stands for an extension of fields/division rings with involution. On top, the prime field is written down (without involution). In the end, we 
want to end up with a GQT $\mathsf{GQT}(\ell,\sigma_\ell)$ which induces $\mathsf{GQT}(k,\sigma_k)$ and $\mathsf{GQT}(k',\sigma_{k'})$.}
\end{figure}

\subsection*{SCHNOR'S RESULT ON INVOLUTIONS}

The bad news is that a general comparison dream cannot hold, as was shown in \cite{GQT}. However, the result is true if we suppose $k$ to {\em be algebraically closed to begin with}, by a result of Schnor \cite{Schnor} (which says that if $\ell \Big/ k$ is a field extension of algebraically closed fields, and $\gamma$ is an involution of $k$, then there exists an involution $\overline{\gamma}$ of $\ell$ which fixes $k$ and induces $\gamma$ in $k$).

\begin{theorem}[Embedding Theorem of quantum theories \cite{GQT}]
\label{emb}
Any GQT over an algebraically closed field $k$ with involution $\gamma$ is embeddable in a GQT over $\ell$, where $\ell$ is any algebraically closed field extension 
of $k$.
\end{theorem}

In particular, this is also true for AQT: \ul{we can embed AQT in an infinite number of ``nonisomorphic'' GQTs over algebraically closed fields in characteristic $0$, and this aspect adds an infinite number of layers to the theory which can be used in various situations (such as quantum coding schemes).}

\medskip
\subsection{The minimal model: $\overline{\mathbb{Q}}$}

Recall the following basic result: 

\begin{theorem}
\label{count}
Let $k$ be any field. If $k$ is not finite,  then $\vert \overline{k} \vert = \vert k \vert$, where $\overline{k}$ is an algebraic closure of $k$; if $k$ is finite, $\overline{k}$ is countable.
\end{theorem}

It's important to note that the statement relies on the Axiom of Choice!

Since $\mathbb{Q}$ is the unique prime field in characteristic $0$, each field of characteristic $0$ contains $\mathbb{Q}$, and hence each algebraically closed 
field $k$ in characteristic $0$ contains the algebraically closed field $\overline{\mathbb{Q}}$ as well.  By Theorem \ref{count}, $\overline{\mathbb{Q}}$ is countable, and hence it is also minimal with respect to being algebraically closed.  

\begin{observation}
The set of GQTs over $\overline{\mathbb{Q}}$ can be considered as the set of \ul{minimal models (over algebraically closed fields) in characteristic $0$}.  
\end{observation}

The idea is that by Theorem \ref{emb}, we know that any minimal GQT can be 
embedded in a GQT over any given algebraically closed field in characteristic $0$. And also, if $k$ is algebraically closed in characteristic $0$ and we consider a GQT over $k$ with involution $\sigma$, one observes that $\sigma$ fixes the prime field $\mathbb{Q}$, and so it also fixes the algebraic closure $\overline{\mathbb{Q}}$.\footnote{Note that the induced action might be trivial, in which case the induced quantum theory over $\overline{\mathbb{Q}}$ is orthogonal/symmetric.}  

In fact, we can do a bit better:

\begin{observation}
Each general quantum theory $\mT$ over an algebraically closed field $k$ in characteristic $0$ induces a general quantum theory over $\overline{\mathbb{Q}}$. 
\end{observation}

{\em Proof}.\quad
Suppose $\sigma$ is the involutory automorphism which comes with $\mT$. As we have seen, it fixes $\overline{\mathbb{Q}}$. If $\sigma$ would fix $\overline{\mathbb{Q}}$ elementwise, then we obtain a nontrivial element of order $2$ in $\mathsf{Gal}(\C/\overline{\mathbb{Q}})$ (where the latter denotes the automorphism group of $\C$ which fixes its subfield $\overline{\mathbb{Q}}$ elementwise), which is a contradiction since this Galois group is known to be torsion-free (that is, contains no nontrivial elements of finite order). \eop \\

\begin{tcolorbox}
\begin{equation}
\mbox{GQT over $\overline{\mathbb{Q}}$}\ \ \ \ \ {\scalebox{2}{\(\rightleftarrows\)}}\ \ \ \ \  \mbox{GQT over $k$}
\end{equation}
\end{tcolorbox}

From this important point of view, it feels more natural to consider quantum theory coordinatized by the algebraic closure $\overline{\mathbb{Q}}$ of the rational numbers. In fact, as 
the rational numbers are dense in the real numbers, the expression 
\begin{equation}
\mathbb{Q} = \mathbb{Q} + i\mathbb{Q} \ \subset\ \overline{\mathbb{Q}} + i\overline{\mathbb{Q}} \ \subset \ \R + i\R \ =\ \C
\end{equation}   
shows that every element of $\C$ can be seen as a limit of Cauchy sequences in $\overline{\mathbb{Q}}$, in other words: 

\begin{observation}[Universality of the minimal model]
Classical quantum theory can be perfectly approximated by modal quantum theory over $\overline{\mathbb{Q}}$, while the latter is countable, also algebraically closed and contained in every division ring (and in particular: field) in characteristic $0$.  \eop
\end{observation}

\subsection{Finite fields}

In \cite{MQT}, Schumacher and Westmoreland introduced modal quantum theory (MQT) as a finite ``toy model'' for AQT, in which the underlying field $\mathbb{C}$ is replaced by
a finite field $\mathbb{F}_q$ (where $q$ is an arbitrary prime power). Inner products in the usual formal sense are not defined on vector spaces over a finite field, and hence this aspect is not covered in \cite{MQT}.  This means that the very notion of ``orthogonal states'' does not occur in their approach. In  \cite{Lev}, vector spaces are considered over finite fields $\mathbb{F}_p$ with $p$ a prime, for which the following property holds: 
\begin{quote}
$-1$ is not a square in $\mathbb{F}_p$, but it is in $\mathbb{F}_{p^2}$. 
\end{quote}

The reason is obvious: besides the many similarities between $\Big(\mathbb{C},\mathbb{R}\Big)$ and $\Big(\F_{p^2},\F_p\Big)$, one disposes of a natural Hermitian bilinear form which shares important aspects with the inner product $\Big\langle \cdot,\cdot \Big\rangle$. In \cite{GQT} we showed that there is no need at all for restricting the theory to primes with the aforementioned property. Here is a quick overview. 

\begin{itemize}
\item
Let $q$ be any prime power; then up to isomorphism $\mathbb{F}_q$ has a unique extension of degree $2$, namely $\mathbb{F}_{q^2}$. The map
\begin{equation}
\gamma: \mathbb{F}_{q^2} \mapsto \mathbb{F}_{q^2}: a \mapsto a^q
\end{equation}
sends each element of $\mathbb{F}_q$ to itself, while being an involutory automorphism of $\mathbb{F}_{q^2}$. 

\item
Let $n$ be any positive integer different from $0$; then if $V = V(n,q^2)$ is the $n$-dimensional vector space over $\mathbb{F}_{q^2}$, define for $x = (x_1,\ldots,x_n)$ and $y = (y_1,\ldots,y_n)$ in $V$, 
\begin{equation}
\Big\langle x,y \Big\rangle := x_1^{\gamma}y_1 + \cdots + x_n^{\gamma}y_n. 
\end{equation}

\item
For $\rho, \rho' \in \mathbb{F}_{q^2}$ we have that 
\begin{equation}
\Big\langle \rho x,\rho' y\Big\rangle = \rho^{\gamma}\Big\langle x,y \Big\rangle \rho',\ \mbox{and}\ \Big\langle x,y\Big\rangle^{\gamma} = \Big\langle y,x\Big\rangle.
\end{equation}
\end{itemize}

The following observation is taken from \cite{GQT}.

\begin{observation}
The linear $(n \times n)$-matrices $U$ which preserve the form $\langle \cdot,\cdot \rangle$ precisely are unitary matrices: $(n \times n)$-matrices $U$ for which $U^* U$ is the $(n \times n)$-identity matrix, where $U^* := {(U^{\gamma})}^T$. 
\end{observation}

Classical quantum theory compares with MQT over finite fields as follows.  

\begin{tcolorbox} 
\begin{remark}[$(\C,\R)$-Analogy | finite fields version] {\rm 
In this model of QT, $\mathbb{F}_{q^2}$ plays the role of $\mathbb{C}$, $\mathbb{F}_q$ the role of $\mathbb{R}$, $\gamma$ the role of complex conjugation, and $\langle \cdot,\cdot \rangle$ the role of inner product. By choosing any element $\kappa$ in $\mathbb{F}_{q^2} \setminus \F_q$, we can represent each element of $\mathbb{F}_{q^2}$ uniquely as 
\begin{equation}
a + \kappa b,
\end{equation}
with $a, b \in \F_q$. So viewed from this representation, the situation at least looks ``a little classical.''}
\end{remark} 
\end{tcolorbox}

\subsection{The base field in fixed characteristic: quantum Lefschetz}
\label{ql1}

If we agree that in the K\o benhavn interpretation we need an algebraically closed field $\ell$ for describing a. o. a theory of observables based on eigenvalues and eigenvectors, we still need to decide in which {\em characteristic} it lives. Once a characteristic $p \geq 0$ is fixed, another fundamental question emerges: 
\begin{quote}
Which base field do we select among the division rings of characteristic $p$? 
\end{quote}
The very question as of which base field is needed in the K\o benhaving interpretation has a long history, and has been considered in many papers. Here is a very short overview of a number of authors which consider the ``base field question'' in recent work, as we compare it to our own GQT-approach in \cite{GQT}.
\begin{itemize}
{\footnotesize 
\item[{\bf BARRET/HARDY}]

\quad\ {\Large {\color{orange} $\odot$}} Both Hardy and Barret see states as probability vectors in some vector space $V$, and  
as the probability entries are real numbers (contained in the interval $(0, 1)$), $V$ is assumed to be a real vector space \cite{GPT,Hardy}. This means that the underlying 
algebraic structure is assumed to contain the field of real numbers.  

In the unifying viewpoint of GQTs \cite{GQT}, probabilities are manifestations 
of the Hermitian form (through the generalized Born rule, e.g.), and the field or division ring one uses as underlying algebraic structure (whatever it is). 

{\Large {\color{orange} $\odot$}} In \cite{Hardy}, two integer parameters $K$ and $N$ emerge for which the identity $K = N^2$ holds. If one considers underlying 
algebraic structures such as the real numbers $\mathbb{R}$, the complex numbers $\C$ or the quaternions $\mathbb{H}$, only $\C$ confirms the aforementioned identity. Hardy concludes that this | at least  intuitively | points towards 
the complex numbers, without providing a formal proof \cite{Hardypriv}. But the identity $K = N^2$ was not considered in the entire realm of fields and division rings in characteristic $0$ | only over the set $\{ \C, \mathbb{R}, \mathbb{H} \}$. (On the other hand, assuming the probabilities to be rational numbers in $(0,1)$ would also yield more flexibility for $V$.)
Barret's generalized probabilistic theories are based on Hardy's axiomatic approach, so he ends up with the complex numbers as well.   

In our approach of GQTs \cite{GQT}, we also work with vector spaces, but any division ring (with involution) is allowed to be 
the coordinatizing agent, so as to find unifying behavior in this universe of quantum theories.  The no-cloning result of \cite{GQT}, for instance, solely follows from the concept of linearity/superposition and works for all division rings | hence also fields and algebraically closed fields, such as in particular $\C$. It shows that no-cloning is not a particular instance at all of quantum theory represented in the framework of complex numbers. 

\item[{\bf CASSINELLI--LAHTI}]

In \cite{CALA}, a program is outlined for an axiomatic reconstruction of quantum mechanics based on the statistical duality of states and considerations of symmetry. The authors discuss the choice of the complex numbers in \cite[section 4]{CALA}. In their viewpoint of {\em orthomodular spaces}  and because of the use of Sol\'{e}r's theorem,  one is again coordinatizing over a  member of $\{ \mathbb{R}, \mathbb{C}, \mathbb{H}\}$. The authors personally prefer the complex numbers for reasons of elegance. (At this point it is important to recall the dangers of the elegance-aspect | cf. the introduction of this paper!) } 
\end{itemize}

We also refer to \cite{AEDA, BAEZ, CALA2, SOL} (from that paper) for relevant papers. 

The discussion of the current subsection motivates us to introduce a {\em quantum Lefschetz formalism} in the next section. 


\section{Quantum Lefschetz Principle A}
\label{Lefsch}

In this section we introduce the quantum Lefschetz principle. We first explain the {\em mathematical} Lefschetz principle. 

\subsection{Lefschetz principle}

The principle in its most naive but perhaps clearest form states the following:

\begin{quote}
``Every true statement about an algebraic variety over the complex numbers $\C$ is also true for a algebraic variety 
over {\em any} algebraically closed field $\ell$.''  
\end{quote}

In fact, the naive formulation of the main principle is even stronger: to check whether a statement about an algebraic variety over 
an algebraically closed field $\ell$ in characteristic $0$ is true, it is sufficient to check it over $\C$ (whatever that means). Lefschetz, in his book \cite{Lef}, states:
\begin{quote}
``In a certain sense, algebraic geometry over a ground field of characteristic $0$ may be reduced to complex algebraic geometry.''
\end{quote}

(Recall the discussion in the introduction concerning the ``choice of $\C$.'')
As Seidenbergh remarks in \cite{Seiden}, the situation is not quite as simple as Lefschetz claims. He describes the following beautiful yet simple example: consider two curves defined by equations $f(X, Y, Z) = 0$ and $g(X, Y, Z) = 0$ in the projective plane over the algebraically closed field $k$ of characteristic $0$. In the complex case, it is well known that such curves meet in at least one point: there exist 
numbers $a, b, c \in \C$ such that 
\begin{equation}
f(a, b, c)\ =\ 0\ =\ g(a, b, c).
\end{equation}

According to Lefschetz, we could conclude the same for $k$. Indeed, assume that $k$ is the algebraic closure of a field which is finitely generated over $\mathbb{Q}$, and which is contained in $\C$, and hence that $k$ is a subfield of $\C$ (cf. Theorem \ref{baby} below). We then conclude that the curves have a point in common (because they have a point in common over $\C$). Only now, the situation has changed: the property we were set to obtain was that the curves {\em have} a point in common {\em over $k$}, but now, we can only conclude that they have a point in common in some extension field of $k$ inside of $\C$!

So we obviously need more precise statements in order to grasp its depth. 
The essence of the principle becomes much clearer if we first look at the|precise| baby version first: 

\begin{tcolorbox}
\begin{theorem}[Baby Lefschetz Principle]
\label{baby}
Let $k$ be a field of characteristic $0$ which is finitely generated over the field of rationals $\mathbb{Q}$. Then there is an 
isomorphism of fields 
\[ \gamma:\ k \ \mapsto\ \gamma(k)     \]
from $k$ to a subfield $\gamma(k)$ of the field of complex numbers $\C$.
\end{theorem}
\end{tcolorbox}

This statement|although simple| is extremely powerful once we start thinking about its consequences. Lefschetz observed that 
every equation over an algebraically closed field $\ell$ of characteristic $0$ is defined by a finite number of coefficients, which generate a field $\widetilde{\ell}$ which is finitely generated over $\mathbb{Q}$, and whence Theorem \ref{baby} applies. The idea is very deep: although we start in any algebraically closed field of characteristic $0$, the typical problems which occur in the theory of algebraic varieties involve only finite data, and using Theorem \ref{baby} we obtain a principle which {\em transfers} the problem to the complex numbers. Unfortunately, as we have seen, Lefschetz's initial version was not precise. Tarski came up with a solution in \cite{Tarski}:

\begin{tcolorbox}
\begin{theorem}[Minor Lefschetz Principle]
\label{minor}
The theory of algebraically closed fields of characteristic $0$ admits quantifier elimination, and therefore all models 
are elementary equivalent. 
\end{theorem}
\end{tcolorbox}

To be clear, we also provide the following equivalent formulation.

\begin{tcolorbox}
\begin{theorem}[Minor Lefschetz Principle, version 2]
\label{minor2}
If an elementary sentence holds for one algebraically closed field of characteristic $0$, then it holds for {\em every} algebraically closed field of characteristic $0$. 
\end{theorem}
\end{tcolorbox}

We recall the notion of {\em elementary sentence} for the convenience of the reader. Let $\ell$ be any field. An {\em atomic formula} relative to $\ell$ is an expression of the form $f = 0$, where $f$ is a polynomial with coefficients in $\ell$. By a {\em formula} (again relative to $\ell$) we mean an expression built up in a finite number of steps from atomic formulae by means of conjunction, negation and quantifiers of the form ``there exists an $x$ such that,'' where $x$ varies over an algebraically closed field $L$ containing $\ell$. An {\em elementary sentence} (relative to $\ell$) then is a formula involving no free parameters.

Another very interesting variation of Lefschetz's principle is the following.

\begin{tcolorbox}
\begin{theorem}[Algebraic Lefschetz Principles]
\label{ALP}
Let $\Phi$ be a sentence in the language $\mL_r = \{ 0, 1, +, -, \cdot \}$ for rings, where $0, 1$ are constants and $+, -, \cdot$ are binary functions. 
The following are equivalent:
\begin{itemize}
\item
$\Phi$ is true over every algebraically closed field in characteristic $0$; 
\item
$\Phi$ is true over some algebraically closed field in characteristic $0$; 
\item
$\Phi$ is true over algebraically closed fields in characteristic $p \ne 0$ for arbitrarily large primes $p$; 
\item
$\Phi$ is true over algebraically closed fields in characteristic $p \ne 0$ for sufficiently large primes $p$; 
\end{itemize}
\end{theorem}
\end{tcolorbox}

Unfortunately, although the Minor Lefschetz Principle is very general (and even still more general versions are known), not every statement carries over just like that. For example, the statement that the cardinality (number of rational points) of every variety is at most that of the continuum is true over $\C$, but obviously this is not true over algebraically closed fields with greater cardinality than $\C$.

\subsection{Algebraically closed fields in quantum theory}

Even if we agree to work with algebraically closed fields as the base field to describe quantum theory in the K\o benhavn interpretation, why would we have to use the complex numbers? For every prime number $p$ (including $0$) and every cardinal number $\kappa$, there is an algebraically closed field $\ell$ which lives in characteristic $p$ and for which $\vert \ell \vert = \kappa$. And even if the field is not algebraically closed, there are a number of valuable quantum theories around which have much in common with classical quantum theory, but which also still behave differently. For instance, as William Wootters explained to me, quantum theory over the reals contains mysterious features which have not been explained to this day. 
If we write a prepared state $\vert \psi \rangle$ relative to an othonormal eigenbase as $(c_1,c_2,\ldots,c_d)$,  and each $c_k$ as $r_ke^{i\phi_k}$, then only the real vector $(r_1,r_2,\ldots,r_d)$ contains the information about probabilities. Is there an underlying physical reason? 

\begin{remark}
{\rm Suppose we write each $c_k$ as $a_k + ib_k$ ($a_k, b_k$ real). Then all states which ``probability project'' on $(r_1,r_2,\ldots,r_d)$ are precisely of the form $(a_1,b_1,\ldots,a_d,b_d)$ for which $a_k^2 + b_k^2 = r_k^2$ for all $k$ (while the $r_k^2$ sum up to 1). So they are precisely the points of a so-called $2d$-dimensional Clifford torus.} 
\end{remark}


In each of the approaches we have seen in subsection \ref{ql1}, either it is (sometimes implicitly) assumed that the reals are contained in $\ell$, or a number of properties are assumed which in the end hold for a K\o benhavn theory over the complex numbers (but not necessarily characterize the field uniquely as the complex numbers). We propose a totally different approach.  If we start from any given algebraically closed field $\ell$|say, in characteristic $0$ to fix ideas|then maybe a reasoning similar to that of Lefschetz might enable us to transfer the entire quantum theory described in the language of the K\o benhavn interpretation over $\ell$, to the K\o benhavn interpretation over $\C$. So a sufficiently subtle Lefschetz theory in the spirit of the previous theorems might give us an answer. 

\begin{question}
Can we develop a Lefschetz principle for quantum theory, so as to show that one can indeed consider the complex numbers as a suitable field of coordinates?
\end{question}

An answer would largely settle the base field question in quantum theory. For instance, \ul{how much of complex quantum theory can be described in first order logic over $\C$} (plus some appropriate induction principle)? 

Currently we are developing an answer to this fundamental question.



\section{Automorphisms of $\C$ and codes}
\label{autcode}

The most commonly used automorphism of $\C$ in quantum theory is complex conjugation, but there are many others in all models of Zermelo-Fraenkel set theory plus AC. In fact, so many that the structure of $\Aut(\C)$ is very hard to understand. On the other hand, in models without AC, it is consistent to say that $\vert \Aut(\C) \vert = 2$ | that is, that standard complex conjugation is the only nontrivial automorphism of $\C$! Strangely enough, the immense size and complexity of $\Aut(\C)$ (upon accepting AC) is virtually never used in quantum theory, while for instance in quantum coding theory it would be a powerful tool. But even if $\vert \Aut(\C) \vert = 2$, nice things can happen. The projective semi-linear group $\mathsf{P\Gamma L}_N(\C)$ acts on the ($(N - 1)$-dimensional) state space $\mathsf{PG}(N - 1,\C)$, and can select states based on the occurrence of fixed points; if $\vert \Aut(\C) \vert = 2$, one can understand and control the action much better in this context. If we work in such a model, it is easy to show that for each automorphism $\varphi$ of the state space (that is, each element of $\mathsf{P\Gamma L}_N(\C)$), we have that at least one of $\varphi$ or $\varphi^2$ is an element of the projective general linear group, and so has fixed points as the field  $\C$ is algebraically closed. This is a very interesting property in the context of selection processes, and hence also of quantum codes. We will come back to these codes in a future paper \cite{ACcodes}.

\section{Eigenvalues, eigenvectors and probabilities}
\label{eigen}

We start this section with a construction taken from Brunner et al. \cite{BSB}, of weird vector spaces upon not accepting AC. 

\subsection{A particular example from Brunner et al.}

Let $S$ be a set. Then $\ell_2(S)$ is defined as 
\begin{equation}
\ell_2(S)\ =\ \{ x \in \C^S\ \vert\ \parallel x \parallel_2 < \infty \},
\end{equation}
where $\parallel x \parallel_2 = \mathrm{sup}_{E \subseteq S\  \mbox{{\small finite}}} \sqrt{\sum_{s \in E}\vert x(s) \vert^2}$, and where we have identified $x \in \C^S$ with a map $x: S \mapsto \C$.   

Now let $\mR := \{ P_n = \{ a_n, b_n \}\ \vert\ n \in \omega \}$ be a collection of Russel's socks (upon not accepting AC). (In fact, we assume a stronger 
version of Russel's socks, as in \S 1.1 of \cite{BSB}.) Let $\Omega := \cup_{n \in \omega}P_n$ be the set of all socks, and define 
\begin{equation}
\mL\ :=\ \{ x \in \ell_2(\Omega)\ \vert\ (\forall n \in \omega)\ x(a_n) = -x(b_n) \}. 
\end{equation}

In \cite{BSB} it is shown that $\mL$ is an irreflexive complex Hilbert space, so that an operator on $\mL$ cannot be equal to its adjoint and 
the usual Hilbert space formalism of quantum theory fails to work. Brunner, Svozil and Baaz argue in \cite{BSB} that such Hilbert spaces have to be taken into account, through the following thought experiment.

\subsection*{Identical particles} 

Before describing the thought experiment of \cite{BSB}, we recall some theory about identical particles. 

We say that two particles are {\em identical} if all their intrinsic properties such as mass, spin, charge, etc. are exactly the same. The configuration space of $N$ identical particles is defined by 
\begin{equation}
\mC(d,N)\ := \ \Big(\times_N\mathbb{R}^d \setminus \Delta \Big)\Big/\mathrm{Sym}(N).
\end{equation}

Here, the particles live in $\mathbb{R}^d$, $\times_N\mathbb{R}^n$ denotes the cartesian product $\underbrace{\mathbb{R}^n \times \cdots \times \mathbb{R}^n}_{\mbox{$N$ times}}$; $\Delta$ is the subspace of points for which at least two ``coordinates'' (in {$(\mathbb{R}^d)^N$) are the same ({\em mathematical explanation}: to remove singularities; {\em physical explanation}: identical particles cannot occupy the same ``location'' in space, in the sense that no projection on coordinate axes may coincide); and finally, $\mathrm{Sym}(N)$ is the symmetric group on $N$ letters (which has to be divided out since we cannot distinguish between the particles). The fundamental group of $\mC(d,N)$ gives information about the continuous trajectories between the particles. \\

{\em Example}.\quad
Let $d = 1$ and $N = 2$ (two particles moving on a line). Then $\mC(1,2)$ is homeomorphic to the space $\Big(\mathbb{R} \times \mathbb{R} \setminus \Delta \Big)\Big/ \mathrm{Sym}(2)$, in which particles 
$(u,v)$ and $(v,u)$ are identified, and where $\Delta$ is defined by the line $u = v$. So we obtain the half-plane defined by $u > v$. \\ 

Finally, a particle that follows Fermi-Dirac statistics is called a {\em fermion}; generally such particles have a half-odd integer spin.

\subsection*{Thought experiment}

View $\{ a_n, b_n \}$ as an assembly of identical noninteracting spin-$\frac{1}{2}$ particles which obey the Fermi-Dirac statistics. 
Its Hilbert space 
$\mH_n$ is defined as 
\begin{equation}
\mH_n\ :=\ \Big\langle e_1(a_n) \otimes e_2(b_n) - e_2(a_n) \otimes e_1(b_n) \Big\rangle,  
\end{equation} 
and is isomorphic to $\mL_n = \{ x \in \ell_2\Big(\{ a_n, b_n \} \Big)\ \vert\ x(a_n) + a(b_n) = 0 \}$. The family of all socks is viewed as the compound 
system of the distinguishable assemblies. Their Fock space is 
\begin{equation}
\mF\ =\ \oplus_{N \in \omega}\Big( \otimes_{n \in N} \mH_n \Big). 
\end{equation}

The space $\mF$ and $\mL = \oplus_{n}\mL_n$ are counter examples to several assertions in Hilbert space theory \cite{BSB}:
\begin{itemize}
\item
both spaces do not admit an infinite orthonormal (Schauder, cf. the next subsection) eigenbase, so there is no way  to choose a mode of observation (in the sense 
of Bohr's complementarity  interpretation); there is no Hamel base either;   
\item
as the vector space duals of $\mF$ and $\mL$ are different from $\mF$ and $\mL$, there is no notion of self-adjoint operator in both $\mF$ and $\mL$. 
\end{itemize}

\subsection{Schauder bases}

In infinite-dimensional Hilbert spaces, quantum theorists usually work with Schauder bases instead of Hamel bases. Hamel bases are the usual bases considered in linear algebra, but Schauder bases are altogether somewhat different. We say that $\mB$ is a {\em Schauder basis} of the infinite-dimensional Hilbert space $\mH$, if each vector can be represented as a tuple in $\mB$ with at most a countable number of nonzero coefficients. (Each vector can be obtained as a convergent series of vectors generated by $\mB$ seen as a Hamel base of a subspace of $\mH$, so that the subspace generated by $\mB$ as a Hamel base is dense in $\mH$.) We say that a Hilbert space is {\em separable} if it contains a countable Schauder basis. 
It can be shown that all infinite-dimensional separable Hilbert spaces are isometrically isomorphic to $\ell^2$. Note also that by Baire categoricity, one can show that the Hamel dimension of a complex Hilbert space is always finite or uncountable! Here, ``Hilbert'' is important. Suppose $(e_i)$ is an orthonormal basis of $\mH$ and let ${\{ e_{j(i)}\}}_{i \in \mathbb{N}}$ be a subset. Then $\sum_{n = 1}^\infty \frac{1}{n}e_{j(n)} \in \mH$, but it is not expressible relative to $(e_i)$ as a Hamel basis. 

In spaces $\mH$ with a Schauder base $\mB$ Born's probability formalism works perfectly. Note that if $\mH$ is an infinite-dimensional Hilbert space, the dimension refers to its dimension with respect to a Hamel basis, and that the ``Schauder dimension'' can be different than the usual Hamel dimension.  

If one considers an observable $A$ in an infinite-dimensional Hilbert space, the orthonormal eigenbase of $A$ is also considered to be a Schauder base.


\subsection{Projector operators} 

A Hermitian operator $\bP$ of a Hilbert space $\mH$ is called a {\em projector (operator)} if $\bP^2=  \bP$. (It is a dichotomy operator since it only allows two possible outcomes.) Now consider a Hilbert space ${\mH}$, and let $\bP$ be the observable which projects any vector onto the subspace $A$ of $\mH$. Let $\mB$ be a Schauder eigenbase  of $\bP$ (taken that it exists). Then $\mH = A \oplus B$, where $A$ is generated by the eigenvectors in $A$ and $B$ is generated by the eigenvectors in $\mB \setminus A$. Let a quantum system be prepared in the state $\vert \Psi \rangle$. Upon projecting $\vert \Psi \rangle$ onto $A$, respectively $B$, we obtain 
vectors $\vert \Psi \rangle_A$, respectively $\vert \Psi \rangle_B$, and we can write
\begin{equation}
\vert \Psi \rangle\ = \ \vert \Psi \rangle_A\ +\ \vert \Psi \rangle_B.
\end{equation}
Now expand $\vert \Psi \rangle_A$, respectively $\vert \Psi \rangle_B$, in the Schauder base $\mB_A$ of $A$, respectively $\mB_B$ of $\mB$, induced by $\mB$ to 
obtain $\vert \Psi \rangle_A = \sum_{\vert \Psi_i \rangle \in \mB_A}a_i\vert \Psi_i \rangle$ and $\vert \Psi \rangle_B = \sum_{\vert \Psi_i \rangle \in \mB_B}b_i\vert \Psi_i \rangle$. Then the probability $P_A$ to measure the eigenvalue $\lambda = 1$ (``YES'') and  the probability of measuring $\lambda = 0$ (``NO'') are given by 
\begin{equation}
P_A\ :=\ \sum_{\vert \Psi_i \rangle \in \mB_A}\vert a_i \vert^2, \ \ \
P_B\ :=\ \sum_{\vert \Psi_i \rangle \in \mB_B}\vert b_i \vert^2. 
\end{equation}

Note that if $\bP_i$ is the projector operator onto the eigenvector $\vert \Psi_i \rangle \in \mB_A$, then $\bP$ can be easily described as 
\begin{equation}
\bP\ = \ \sum_{\vert \Psi_i \rangle \in \mB_A}\vert \Psi_i \rangle \langle \Psi_i \vert \ =\ \sum_{\vert \Psi_i\rangle \in \mB_A}\bP_i.
\end{equation}


\subsection{Double slit experiments: a variation in quantum theory without AC}

Young's double slit experiment does not need an introduction: a light beam is fired in a straight line through a panel with two disjoint rectangular slits on a screen. Classical Physics would expect a pattern which corresponds to the size and shape of the slits, but that is not what happens. Instead, an interference pattern occurs. This even happens when the experiment involves single particles: through a double split apparatus, particles are sent one at a time and the interference pattern emerges eventually as well. Although the particles are measured 
as a single pulse in a single position, a probability wave describes the probability of observing the particle at a specific point $(x,y)$ in the plane of the screen. Born's rule gives us the probability distribution for finding an electron at specific places of the screen. Once an electron hits the screen and is detected, the wave function collapses and the outcome can be described through the eigenvalues of an observable matrix. \\


Although the following thought experiment is not directly related to the double slit experiment, it still shares some of it (weird) characteristics.

So let $\mH$ be a (generalized) Hilbert space over some field $k$ in a model of Zermelo-Fraenkel without AC, which allows (Schauder) bases $\mB_1$ and $\mB_2$ of different cardinalities. This field may not be the complex numbers, but in the context of modal quantum theories, 
we should still consider this possibility. In any case we know that $\mH$ exist by \cite{Laugh}. Note that $\mH$ necessarily is infinite-dimensional (in the sense that it is not finite-dimensional). 
Now let $\widetilde{\mH}$ be a second Hilbert space over $k$, and 
consider $\widehat{\mH} := \mH \oplus \widetilde{\mH}$. Let $\bP_\mH$ be the projector operator of $\widehat{\mH}$ onto $\mH$; this is an observable with eigenvalues $0$ and $1$, and acts on $\mH$ as the identity. Now consider a state $\vert \Psi \rangle$ in $\widehat{H}$. After measuring $\bP_{\mH}$, $\vert \Psi \rangle$ collapses into a state in $\mH$ which is a superposition. 

One question now is: 

\begin{question}
What is the probability that a state is contained in $\mH$? As $\mH$ has no well-defined dimension, we have no idea how ``big'' $\mH$ is with respect to $\widehat{\mH}$.
\end{question}

But we can do better. 

\subsection*{Thought experiment: AC black box measurements}

Suppose a quantum system is prepared in a state $\vert \Psi \rangle$ in the (possibly generalized) Hilbert space $\mH$ over the field $k$. We assume that upon not accepting the Axiom of Choice, $\mH$ admits Schauder bases of different cardinalities. Now we are going to perform a measurement corresponding to the Hermitian operator $A$. Before making the measurement, we \ul{do not know whether the Axiom of Choice holds true or not}; this can only  --- in principle --- be observed after the measurement has been made. (There is a black box which returns the value ``$0$'' or ``$1$.'') After the measurement, we obtain an eigenvalue $\lambda$ with probability $p_\lambda$. If AC holds in the underlying mathematical theory, this outcome is standard. If AC would not hold, $\lambda$ could have been measured with a different probability (by the formalism below, for instance, $p_\lambda \ne 0$ could be infinitely small). \\

{\footnotesize
\begin{quote}
The analogy with the double slit experiment goes as follows. The case in which AC is true (the black box returns $1$), and a classical measurement is performed, compares to the classical observation of two slits.  The case in which AC is not true (the black box returns $0$), and a weird measurement is performed, compares to the observation of interference.  

\end{quote}
}


\subsection{New Born formalism and higher Schauder bases}

Let $S$ be an uncountable set, and suppose $\mC := \{ r_s\ \vert\ s \in S \}$ is a set of positive real numbers. Suppose $\sum_{s \in S}r_s = R$ as a limit is also real. Then in any model of Zermelo-Fraenkel with AC, it can be shown that only at most a countable number of elements in $\mC$ are different from zero. The proof uses the fact that a countable union of finite sets is also countable | a fact which fails miserably when AC is not around. So it makes sense to define {\em higher Schauder bases} as follows.  We say that $\mB$ is a {\em higher Schauder basis} of the infinite-dimensional Hilbert space $\mH$ if all vectors of $\mH$ can be 
represented by a unique $\vert \mB \vert$-tuple in which an uncountable number of nonzero entries is allowed and so that such vectors also occur. It follows that $\mB$ is not countable.   
Consider a state $\vert \Psi \rangle  = {( a_b )}_{b \in \mB}$. For Born's formalism to work, we want 
\begin{equation}
\label{eqbo}
\sum_{b \in \mB}\vert a_b \vert^2 \ =\ 1.
\end{equation} 
Upon accepting the Axiom of Choice it  is easy to show that the latter expression implies that only a countable number of entries in $ {( a_b )}_{b \in \mB}$ is nonzero, and so in $\mH$ we can still consider those states which makes sense in the quantum-theoretical setting. If we work in Zermelo-Fraenkel set theory without choice however, there are models in which (\ref{eqbo}) is true, with an uncountable number of entries nonzero \cite{MOsum}! In this formalism, we only consider state vectors $\vert \Psi \rangle  = {( a_b )}_{b \in \mB}$ for which $\sum_{b \in \mB}\vert a_b \vert^2 \in \R$ (before normalization). (If one would work over an algebraically closed field $\ell$ of characteristic $0$ which is different from $\C$, then we 
ask that $\sum_{b \in \mB}\vert a_b \vert^2$ is contained in the real-closed subfield which is defined relative to the choice of complex conjugation.)

\begin{question}
Does this formalism cover quantum-theoretical situations which are not possible over classical Schauder bases?
\end{question}

We suspect that the answer of this question, and also of the next question is ``yes.''

\begin{question}
Does it make sense to introduce \ul{nonstandard} probabilities (e. g., infinitely small probabilities) in this context?\\
\end{question}

\subsection{Blass's Theorem and Ineffective observables}

In \cite{Blass}, Blass showed in Zermelo-Fraenkel set theory  that if every vector space has a basis, then AC also holds. He starts with a set $\mX$ of disjoint nonempty sets $X_i$ ($i \in I$), picks an arbitrary field $k$, and constructs the field extension $k(X)$, where $X = \cup_{X_i \in \mX}X_i$. Then he constructs a particular subfield $K$ of $k(X)$, and interprets $k(X)$ as a vector space over $K$. In $k(X)$ he considers the $K$-subspace $V$ spanned by $X$. 
Assuming that $V$ has a basis, he then constructs a choice function on $\mX$. Unfortunately, it does not follow that AC is deducible from the statement that {\em every} $\ell$-vector space has a basis, with $\ell$ a specified, fixed field. On the other hand, obviously $k$, $k(X)$ and $K$ live in the same characteristic, so we have the following stronger statement.

 \begin{tcolorbox}
\begin{theorem}[Blass's Theorem, version 2]
\label{Blass}
Let $p$ be any prime, or $0$. Then in Zermelo-Fraenkel set theory, AC is deducible from the assertion that every vector space over a field of characteristic $p$ has a basis. 
\end{theorem}
\end{tcolorbox}

For quantum theory the importance is obvious: in the classical K\o benhavn formalism, observables (Hermitian operators) collapse into one of the vectors of an orthogonal eigenbase, and the corresponding eigenvalue is the resulting observed value. Upon not accepting  AC and working in models of ZF-theory without AC, it could very well happen that some Hilbert spaces $\mH$  over $k = \C$ (or some other field) do not have a base, so that the formalism of Hermitian observables fails, or needs to be adapted at the very least. 
In any case, by Theorem \ref{Blass}, we may suppose that the characteristic of $k$ is $0$.  
For instance, let $\mB$ be the set of orthonormal eigenvectors of some given observable $B$ (which cannot be maximal by assumption), and let $\langle \mB \rangle$ be the subspace of $\mH$ generated by $\mB$ over $k$ (either as a Hamel base or as a Schauder base). By taking a state $\vert \Psi \rangle$ outside of $\langle \mB \rangle$, we cannot perform a measurement using the state $\vert \Psi \rangle$.

\begin{remark}[Quantum Lefschetz Principle B]
{\rm Is $\C$ a candidate? If not, in view of first order logic we may switch to an other algebraically closed field for 
which Blass's Theorem does work (as such ending up with ineffective observables). }
\end{remark}


\end{document}